\documentstyle[aps,prl,graphicx,amssymb,multicol,times]{revtex}
\draft
\newcommand{\blackbox}{\rule{2mm}{2mm}}
\newcommand{\smbbox}{\rule{1.7mm}{1.7mm}}
\begin{document}
\title{Levitation of Current Carrying States in the Lattice Model for
the Integer Quantum Hall Effect}
\author{Th. Koschny, H. Potempa, and L. Schweitzer}
\address{Physikalisch-Technische Bundesanstalt, Bundesallee 100,
38116 Braunschweig, Germany}
\date{\today}
\maketitle
\begin{abstract}
The disorder driven quantum Hall to insulator transition is investigated 
for a two-dimensional lattice model. The Hall conductivity and the
localization length are calculated numerically near the transition.
For uncorrelated and weakly correlated disorder potentials the current
carrying states are annihilated by the negative Chern states
originating from the band center. 
In the presence of correlated disorder potentials with correlation length
larger than approximately half the lattice constant the floating up of
the critical states in energy without merging is observed. 
This behavior is similar to the levitation scenario proposed for the 
continuum model.
\end{abstract}


\begin{multicols}{2}
Soon after the discovery of the integer quantum Hall effect (QHE)
%\cite{KDP80}
the fate of the current carrying states had been of much
concern. At zero temperature and magnetic field $B=0$ a disordered 
two-dimensional electron gas should behave as an Anderson insulator. 
Therefore the question arose, how the transition from the QHE state 
to the insulating phase takes place when $B$ goes to zero. 
Alternatively, for fixed magnetic field one can also consider the 
increase of the disorder and the broadening of the Landau bands.   
In the presence of disorder, both Khmelnitzkii \cite{Khm84} and Laughlin 
\cite{Lau84} argued that the current carrying states that reside at
the center of the Landau bands cannot disappear discontinuously as 
$B\to 0$, but rather float up in energy above the Fermi level. 
These general quasi-classical considerations which are based on 
scaling theory and supported by
numerical results \cite{And84KHA95} temporarily settled the case. 
Furthermore, the floating up of the critical energies to infinity
became an essential cornerstone in constructing a global phase diagram (GPD)
\cite{KLZ92} of the QHE. 
The levitation picture also implies that in the disorder\,--\,magnetic
field plane only such transitions between QHE-states are possible at which 
the change of the quantized Hall conductivity, 
$\Delta\sigma_{xy}$, is $\pm e^2/h$. 
Therefore, the final transition to the insulator must occur always from the
lowest ($\nu=1$) Hall plateau. 

Recently, the appealing levitation scenario has been questioned both
experimentally and theoretically and the debate about the fate of the 
current carrying states and the appearance of the phase diagram has
been restarted \cite{SKD93GJJ95,KMFP95,Hea99c,SW99b}. 
One reason for this development is probably the fact that despite 
considerable efforts \cite{SR95GR96YB96HY97Fog98} the microscopic
origin of the levitation remains still unclear. 
Although transitions to the insulator from $\nu=1$, as expected from 
the GPD, and the floating up of the current carrying states 
have been observed in some experiments \cite{SKD93GJJ95,JJWH93,Wea94}, 
direct transitions from higher
Hall plateaus ($\nu > 2$) which according to the GPD should not be possible
have also been reported \cite{KMFP95,Sea97a,LCSL98,Hea99c}. 
On the numerical side, the observation of direct QHE to insulator 
transitions with $\Delta\sigma_{xy} > e^2/h$
in tight-binding lattice models and the absence of floating states  
\cite{LXN96XLSN96,SW97,SW98}
have led these authors to conclude that due to lattice effects which
are always present in real systems the conventional floating up model 
which neglects the periodic background potential is inappropriate for 
explaining the experimental results. In addition, a new universality
class has been assigned to direct transitions from the QH-liquid to
the insulating phase where at the critical point the conductivities are  
found to be equal, $\sigma_{xy}^c=\sigma_{xx}^c$ \cite{Sea97a,SW98}.

In fact, the studies presented in \cite{LXN96XLSN96,SW97,SW98,PBS98,HIM99}
showed that with increasing disorder the critical states do not float
up in energy, but instead, the states with opposite Chern numbers from the
center of the tight-binding band move downwards and finally annihilate the
current carrying states below the Fermi energy. In lattice systems the 
extended states can be characterized by a topological nonzero Chern integer 
%\cite{TKNN82,NTW85,Aea88}
\cite{chern}. Even with next nearest neighbor hopping, 
which breaks the particle-hole symmetry, no floating up of states across 
the Landau gap, but only the moving outward of states with opposite 
Chern numbers, away from the center, has been found \cite{YB99}. 
This holds also for general magnetic fields commensurate with the lattice 
size, which are given by a rational number of flux quanta per plaquette, 
$B=p/q\cdot h/(e a^2)$ with $p\ne 1$, that lead to Chern
numbers for sub-bands which are $\neq \pm 1$ \cite{YB99}. Here
$p$ and $q$ are coprime integers, and $a$ is the lattice constant. 
 
Although successful in explaining the experimental observation of direct
transitions to the insulator from higher Hall plateaus, the lattice model
investigated in \cite{LXN96XLSN96,SW97,SW98,PBS98,HIM99} 
possesses a serious drawback.
If magnetic field and lattice effects are weak as in the experimental
situation in GaAs heterostructures an effective mass approximation 
should be appropriate and allow for the floating up results of
the continuum model. 
In the latter, states with negative Chern numbers are absent so that
the proposed annihilation cannot occur.

The aim of our paper is to resolve this controversy and to reconcile the
predictions for the continuum model with the results of the lattice model.
The key for achieving this goal is to refrain from unphysical realizations
of the disorder potentials usually considered in this context. 
We will show that if the assumption of independent random site
energies is abandoned
and long range correlations of the disorder potentials are allowed 
as it is the case in the experiments \cite{WLTP92}, 
the floating up of critical states
across the Landau gap is also seen in the lattice model.

We consider a single-band tight-binding Hamiltonian describing the 
properties of non-interacting particles in a disordered two-dimensional 
system in the presence of a perpendicular magnetic field,
\begin{equation}\label{AH}
H=\sum_{k}w_k^{}c_{k}^{\dagger}c_k^{}+
\sum_{<kl>}V(e^{ib_{kl}}c_{k}^{\dagger}c_l^{}+
e^{-ib_{kl}}c_{l}^{\dagger}c_k^{}),
\end{equation}
where $c_k$ is the fermionic operator on lattice site $k$. The
magnetic field enters the transfer terms connecting nearest neighbors
via the phases which in the Landau gauge read 
$b_{kl} = \pm 2\pi (p/q) (\vec{r}_k\cdot\vec{e}_y)/a$, if
$\vec{r}_l=\vec{r}_k\pm a\vec{e}_x$, and $b_{kl} = 0$ else, 
where $\vec{r}_k$ is the position of site $k$ and  $\vec{e}_x$, 
$\vec{e}_y$ are unit vectors pointing in the $x$ and $y$ directions.
$W$ is the disorder strength of the correlated random potentials 
$w_k \in [-W,W]$. We choose $V=1$ and $p/q=1/8$ so that the 
tight-binding band splits into 8 sub-bands. 

The behaviour of the current carrying states is investigated through a
numerical calculation of the localization length $\lambda_M$ 
\begin{equation}
\lambda_M^{-1}(E,W)=-\lim_{L\to\infty}\frac{1}{2L}
\ln\textnormal{Tr}|G_{1L}^{+}|^2,
\end{equation}
and the Hall conductivity 
\begin{eqnarray}
\sigma_{xy}(E,W;M) & = &
- \lim_{\varepsilon \rightarrow 0^+}\lim_{L^{\rule{0mm}{1.2mm}} 
\rightarrow \infty}
\frac{e^2}{h}\,\frac{2}{LM}\times\\ 
\text{Tr}\bigg\{\sum_{n}^{L} i \varepsilon (G_{nn}^{+}\nonumber &-&  
G_{nn}^{-}) x_n\, y_n  - 2 \sum_{n,n'}^{L} {\varepsilon}^2 
G_{nn'}^{+} y_n G_{n'n}^{-} x_n   
\bigg\}.
\end{eqnarray}
The $G_{mn}^{\pm}$ are $M\!\times\!M$ sub-matrices of the advanced 
and retarded 
Green functions, $G^{\pm}=(E-H\pm i\varepsilon)^{-1}$, acting in the 
subspace of the columns $m$, $n$ on the lattice. The $x_n$ and $y_n$ are 
$M\!\times\!M$ matrices with elements $(x_n)_{ij}=n\delta_{ij}$ and
$(y_n)_{ij}=j\delta_{ij}$.
Periodic boundary conditions are applied in the y-direction
for the calculation of $\lambda_M$ while $\sigma_{xy}$ is obtained for 
Dirichlet boundary conditions. The length ($x$-direction) and width
($y$-direction) are denoted by $L$ and $M$, respectively, 
and $\varepsilon$ takes care of the proper thermodynamic limit.

The task is accomplished by means of a recursive Green function method
developed previously \cite{MK83,Mac85}.
This procedure is well suited for calculating very long disordered
2d systems of finite width without getting into trouble with 
storage-space problems. Therefore, 
also the correlated on-site disorder potentials $w_k$ have to 
be generated iteratively site by site. This is implemented in analogy
to the successive creation of Ising spin chains. In this case the 
algorithm can be based on the Gibbs representation of Markov random 
fields \cite{Joh86}. 
The correlated random number $\mu_{x_k,y_k}$ at a given site $k$ 
with coordinates $x_k, y_k$ is calculated with the help of two 
neighboring numbers already determined,
$\mu_{x_k,y_k}=\,(\mu_{x_k-a,y_k}+\mu_{x_k,y_k-a})/2\,+\,2\,e^{-C}\,\gamma$.
The uncorrelated random numbers $\gamma$ are drawn
from an interval $[-1,1]$ with constant probability density.
If $|\mu_{x_k,y_k}|>1$, the value is reflected back into the unit
interval by $\mu_{x_k,y_k}\to \pm 2-\mu_{x_k,y_k}$.
The correlation strength is tuned via the correlation parameter $C$, 
where $C=0$ corresponds to the uncorrelated case.
The correlated random numbers $\mu$ generated in this way are multiplied 
by the disorder strength $W$ which results in the set of disorder potentials 
$w_k\in [-W,W]$ used in Eq.~(\ref{AH}). 

\begin{figure}
\centering
\includegraphics[width=8.65cm]{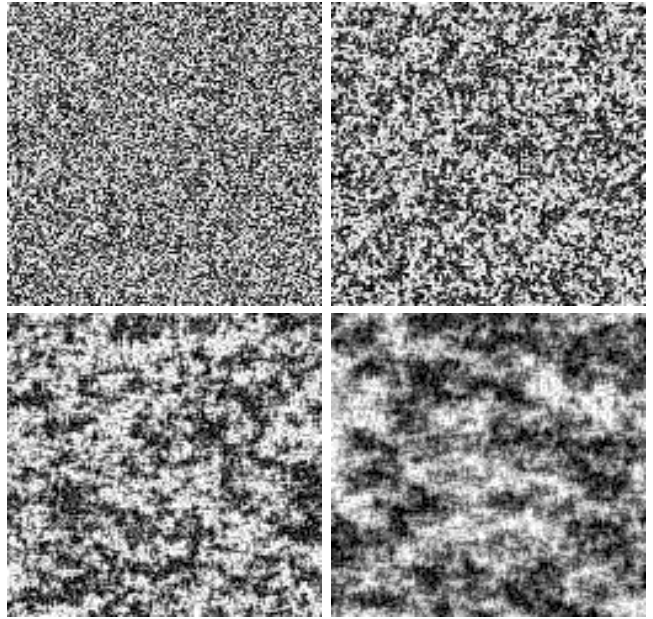}
\smallskip
\caption{Grayscale plot of realizations for a 128 $\times$ 128 array
of correlated disorder potentials with $C=0.0, 0.5, 1.0, 
\text{\ and\ } 1.5$}
\label{corrpot}
\end{figure}

An example is shown in Fig.~\ref{corrpot} where
gray-scale plots for arrays ($128\times 128$) of correlated disorder
potentials with $C= 0.5, 1.0, \textnormal{\ and\ } 1.5$ 
are shown in comparison with the uncorrelated case $C=0.0$ (upper left).
White spots correspond to disorder potentials $w_k\simeq W$ whereas
black ones to $w_k\simeq -W$. 
To better characterize the disorder potentials, we have calculated the 
correlation function  
$K(\rho)=\langle w_k w_l\rangle_{\rho=|\vec{r}_k-\vec{r}_l|}$
averaged over all pairs of disorder potentials at sites
which are a given distance $|\vec{r}_k-\vec{r}_l|$ apart. 
We find an exponential relation, $K(\rho) \sim \exp(-\rho/\eta(C))$, 
where the spatial decay of the correlations is governed by 
the correlation length $\eta(C)$ which for example 
amounts to $\eta(C)/a= 0.8, 1.1, \text{\ and\ } 1.9$ for 
correlation parameters, $C=0.5, 0.7, \text{\ and\ } 1.0$, respectively.  
Thus, for $C<0.7$ the correlation length of the disorder potentials is
less than the magnetic length, $l_B/a=1.1284$ for $p/q=1/8$, while it
is larger for $C\gtrsim 0.7$. 
Our data follow nicely the relation $\eta(C)=1/2\,[\ln(\tanh(C))]^{-1}$
in analogy to Ising spin chains. 

\begin{figure}
\centering
\includegraphics[width=8.5cm]{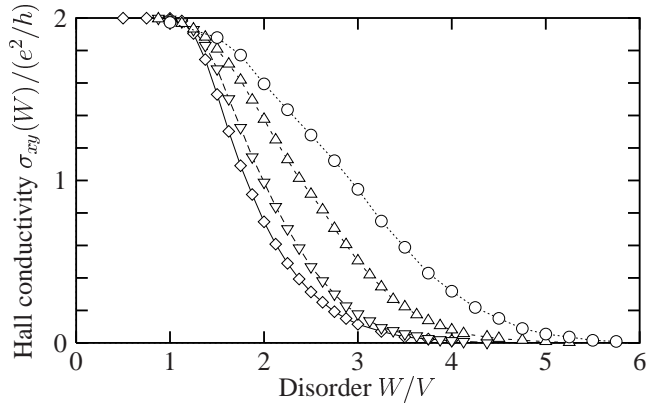}
\smallskip
\caption[]{The Hall conductivity versus disorder at $E=-1.5$ 
(filling factor $\nu\approx 2$). The correlation parameter of the
disorder potentials are $C=0$ 
(\raisebox{-.25mm}{$\Diamond$}), $C=0.5$ ($\triangledown$), 
$C=1.0$ ($\vartriangle$),
and $C=1.5$ (\raisebox{-.4mm}{{\Large$\circ$}}). 
The system length is $L/a=5\cdot 10^{4}$ and the width is
$M/a=128$ ($C=0.0$), $M/a=160$
($C=0.5, 1.0$), and $M/a=256$ ($C=1.5$).}
\label{sigxy}
\end{figure}

First, to trace the QHE to insulator transition we calculate 
the Hall conductivity $\sigma_{xy}(W)$ at energy
$E/V=-1.5$, which corresponds to a filling factor $\nu\approx2$, 
as a function of disorder $W$. The result for various correlation parameters 
is shown in Fig.~\ref{sigxy}. Smooth
transitions from $\sigma_{xy}=2e^2/h$ to zero are observed for correlation 
parameters $C=0, 0.5, 1.0 \textnormal{\ and\ } 1.5$. 
Similar direct transitions have been reported also for larger filling 
factors in the case of uncorrelated disorder potentials 
\cite{SW98,PBS98,SW99b}. This observation and the assignment 
of the critical point to the disorder where 
$\sigma_{xy}=0.5\nu e^2/h=\sigma_{xx}$
have led the authors of \cite{SW98} to conclude that for uncorrelated
disorder potentials this transition corresponds to a new universality
class. 
Although the transition region gets broader with increasing correlations, 
no indication of a Hall plateau at $\sigma_{xy}/(e^2/h)=1$ can be
observed at first sight. From this point of view our
results of the Hall conductivity for correlated disorder potentials
seem qualitatively to be the same as for the uncorrelated case.

\begin{figure}[b]
\centering
\includegraphics[width=8.5cm]{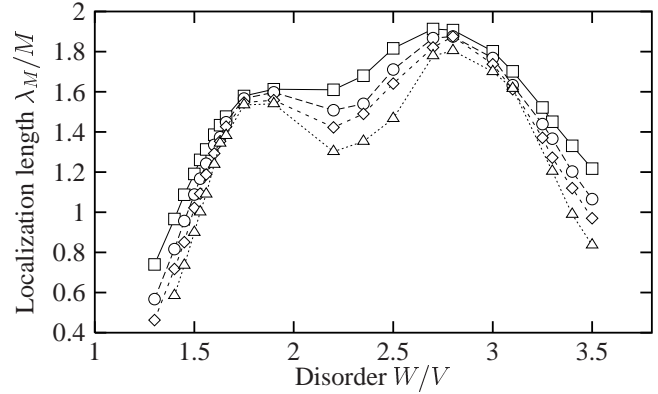}
\smallskip
\caption[]{The normalized localization length versus disorder at
filling factor $\nu=2$. The correlation parameter is $C=1.0$ and the 
system sizes are $M/a=32$ (\raisebox{-.2mm}{$\Box$}), $M/a=64$
(\raisebox{-.4mm}{{\Large$\circ$}}), $M/a=96$
(\raisebox{-.25mm}{$\Diamond$}), and $M/a=160$ ($\vartriangle$).}
\label{Lambda_W}
\end{figure}

Next, to find out the reason for the broadening of the transition region
in the correlated disorder case, we have calculated the
disorder dependence of the localization length.
The position of the current carrying states can be extracted from the
finite size scaling of the normalized localization length $\lambda_M(E,W)/M$.
With increasing system width $M$, $\lambda_M/M$ goes to zero for localized
states while it converges to a finite value for critical states. 
The results for $\nu=2$ and $C=1.0$ are shown in Fig.~\ref{Lambda_W}
where the averaged $\lambda_M/M$ is displayed versus disorder 
for different system widths in the range $32\le M/a \le 160$.  
The averaging is performed over 3 to 5 realizations of the correlated 
disorder potentials and,  for the calculation of $\lambda_M$, $L$ is 
always larger than $5\cdot 10^5$ lattice spacings $a$.
With increasing system width $M$ two distinct
peaks, which correspond to two separate divergences of the localization 
length at about $W_c^{(1)}\approx 1.8$ and $W_c^{(2)}\approx 3.0$, 
emerge out of the broad maximum. 
This implies that two transitions,
from $\sigma_{xy}=2e^2/h \to e^2/h \to 0$, must be hidden in the broad
decay of the Hall conductivity shown in Fig.~\ref{sigxy}.   

In fact, a closer look at the Hall conductivity curves of Fig.~\ref{sigxy}
suggests that there is, hardly noticeable, a tiny shoulder near
$\sigma_{xy}=e^2/h$ for correlation parameters $C=1.0$ and
$C=1.5$. Taking the derivative $d\sigma_{xy}(W)/dW$ 
approximately by $|\Delta \sigma_{xy}(W)/\Delta W|$,
one sees that the peak corresponding to maximal steepness 
is not situated at $\sigma_{xy}=e^2/h$, which should be the case 
for a direct transition, but instead near $\sigma_{xy}/(e^2/h)=1.5$, 
and a second somewhat weaker hump at about $0.5$ can also be detected. 
This clearly means that there is no new single direct transition, 
but two distinct usual transitions which, however, are not easy
to discern in finite size studies. This difficulty of resolving
the quantized Hall steps applies 
also to the experiments in addition to temperature and inelastic 
scattering effects which may mask the transitions.

\begin{figure}
\centering
\includegraphics[width=8.cm]{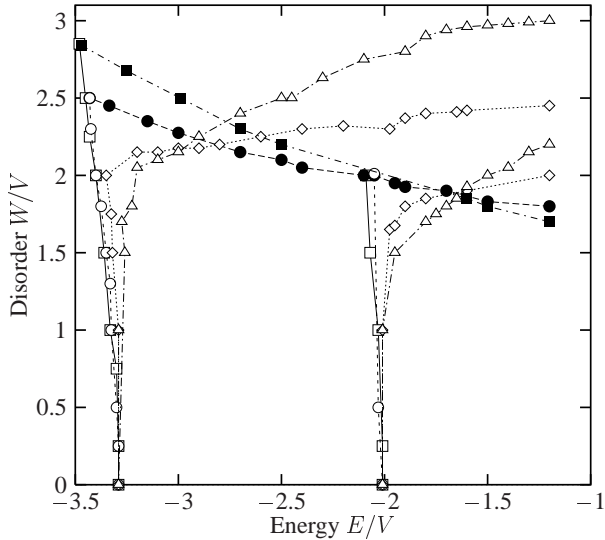}
\smallskip
\caption[]{The position of the critical states in the energy--disorder plane
for various correlation strength of the disorder potentials: 
$C=0.0$ (\raisebox{-.2mm}{$\Box$}, the anti-Chern state is marked by 
\smbbox), $C=0.2$ (\raisebox{-.4mm}{{\Large$\circ$}}, 
the anti-Chern state is \raisebox{-.4mm}{{\Large$\bullet$}}), 
$C=0.5$ (\raisebox{-.25mm}{$\Diamond$}), 
and $C=1.0$ ($\vartriangle$).} 
\label{floating}
\end{figure}

To see the details of the QH to insulator transitions more clearly the
energy and disorder dependence of $\lambda_M(E,W)/M$ has been
calculated in the filling factor range $0 \lesssim \nu \lesssim 2.5 $. 
Depending on the value of the correlation parameter  
we find the surprising result that with increasing disorder the
current carrying states are either annihilated by the anti-Chern
states or do float up in energy as seen in the continuum model.
The essence of our extensive numerical calculations is represented in
Fig.~\ref{floating} which shows the location of the two lowest critical
states within the disorder--energy plane which has been extracted from
the finite size scaling of $\lambda_M/M$. 

We start with looking at the results for the correlation parameter $C=0.2$. 
With increasing disorder $W$ the downward movement of the anti-Chern
states (depicted by {\large$\bullet$}) coming from the band center can be
tracked. The last current carrying state, i.e. that from the lowest
Landau band (marked by \raisebox{-.4mm}{{\Large$\circ$}}) which moves 
slightly to smaller energies, disappears at a disorder 
of about $W/V\approx 2.5$ when both Chern and anti-Chern states touch.
This behavior is similar to the uncorrelated case ($C=0$, Chern and
anti-Chern states are labeled by \raisebox{-.2mm}{$\Box$} and 
\blackbox, respectively) where the last current carrying state 
vanishes at about $W/V\approx 2.85$, also shown in
Fig.\ref{floating}, but hardly to discern from the $C=0.2$ data in
case of the Chern state.

On the other hand, the floating of the energy $E_c$ of the critical 
states across the Landau levels without merging
is seen for both $C=0.5$ and $C=1.0$. Here, the states 
move to higher energies when the disorder is increased. For a given
energy that corresponds to a filling factor in the range $1.5<\nu<2.5$,
the separation between the two critical disorders, which mark the current 
carrying states floating up in energy, is enlarged by enhancing the long
range correlations of the disorder.
The floating up is slower, i.e. a smaller shift in energy is caused by
the same increase in disorder, for larger correlation length. 
For $C=0.5$, the states are closer and float up in energy within a 
small range of disorder values. These data are near a special correlation
strength $C_{s}(B)$ with $\eta\approx a/2$ when the upward floating 
critical states and downward moving anti-Chern states meet halfway.
In that situation the critical disorder $W_c(E)$ is constant which 
implies that only an
infinitesimal change in disorder is needed for the the lowest critical
state to float up and for the anti-Chern state to float down.

In conclusion, we have shown that floating up of current carrying states
with increasing disorder can take place also in the lattice model, 
if correlation in the random disorder potentials is considered. 
In this case, no single direct transition from higher Hall plateaus to
the insulator is seen. Instead, a one by one transition is observed
as it is known from the continuum model. These usual transitions may be,
however, difficult to detect both in experiments and in numerical 
finite size studies. We have shown that \textit{long range} 
correlated disorder potentials will facilitate this task.
Finally, by decreasing the correlation length, a crossover to the known 
annihilation scenario takes place where states with negative Chern number
move downwards and merge with positive Chern states. 
This crossover is found to occur at a correlation length of about half
the lattice constant.

This research is supported by the DFG-Schwerpunktpro\-gramm 
``Quantum-Hall-Systeme'' Grant No. SCHW795/1-1.

\textit{Note added.}--After completing our work we became aware of a
related paper \cite{SWW00} where for Gaussian correlated disorder potentials 
a floating and merging of extended levels was reported. The merging was
not observed in our calculations. 

%\bibliography{qhe,papers2000_database,ludwig}
%\bibliographystyle{prsty}

\end{multicols}

\end{document}